\documentclass[a4paper,twocolumn,11pt,english]{article}
\newcommand{\mathsym}[1]{{}}

\usepackage{amsmath, amssymb, graphics}
\usepackage[T1]{fontenc}
\usepackage[english]{babel}
\usepackage[latin1]{inputenc}
\usepackage{verbatim}
\usepackage{graphicx}
\usepackage{eepic}
\usepackage{latexsym}
\usepackage{color}
\usepackage{indentfirst}
\usepackage{float}
\usepackage{flafter}
\usepackage{afterpage}
\usepackage{makeidx}
\usepackage{authblk}
\usepackage{array}

\makeatletter
\renewcommand\@makefnmark{\@textsuperscript{\normalfont(\@thefnmark)}}
\makeatother
\usepackage[flushmargin]{footmisc}

\usepackage[hang,bf]{caption}
\newcommand{\un}[1]{\mbox{ \rmfamily #1}}
\newcommand{\unp}[1]{\mbox{\rmfamily #1}}

\newcommand{\unrho}[0]{\mbox{ \rmfamily kg} / \mbox{\rmfamily m} ^3}

\newcommand{\url}[1]{\ttfamily #1\normalfont}
\newcommand{\fig}[1]{Fig.~\ref{#1}}
\newcommand{\tab}[1]{Tab.~\ref{#1}}

\newcommand{\app}[1]{App.~\ref{#1}}

\newcommand{\undeg}{\mbox{\textdegree}}

\newcommand{\unsec}{''}

\newcommand{\fo}{Find\_Orb}

\author[1]{Marco Micheli}
\author[1]{David J. Tholen}
\author[1]{Garrett T. Elliott}
\affil[1]{
Institute for Astronomy\\
University of Hawaii\\
2680 Woodlawn Dr., Honolulu, HI 96822\\
USA}

\title{Detection of radiation pressure acting on 2009~BD}
\date{}

\begin{document}

\twocolumn[
\begin{@twocolumnfalse}
\maketitle
\begin{abstract}
{\em
We report the direct detection of radiation pressure on the asteroid 2009~BD, one of the smallest multi-opposition near-Earth objects currently known, with $H \sim 28.4$. Under the purely gravitational model of NEODyS the object is currently considered a possible future impactor, with impact solutions starting in 2071.
The detection of a radiation-related acceleration allows us to estimate an Area to Mass Ratio ($AMR$) for the object, that can be converted (under some assumptions) into a range of possible values for its average density. Our result $AMR=(2.97\pm0.33) \times 10^{-4} \un{m}^2\un{kg}^{-1}$ is compatible with the object being of natural origin, and it is narrow enough to exclude a man-made nature.
The possible origin of this object, its future observability, and the importance of radiation pressure in the impact monitoring process, are also discussed.
}
\end{abstract}
\end{@twocolumnfalse}
]

\section{Introduction}

The search for potentially hazardous asteroids has yielded the discovery of many objects in orbits very similar to the orbit of the Earth. 
Frequently such discoveries trigger the question as to whether the object is of natural or man-made origin, which is usually addressed by integrating the orbit backward in time to see if a close Earth approach occurred at a time consistent with the launch of a rocket into a heliocentric orbit. 

A variety of forces act on solar system objects, including gravity, outgassing, solar radiation pressure, the Yarkovsky effect, the Poynting-Robertson effect, even possibly electrostatic forces. For the vast majority of objects, their motions can be satisfactorily explained by considering only the gravitational force. Even though several of these forces can be acting on an object simultaneously, the contribution from the weaker forces may be undetectable, and therefore no attempt is made to include them when solving for the dynamics of the object. When the observed motion of an object departs from what one would expect from a purely gravitational model, additional forces need to be taken into consideration. Which of these forces gets taken into account can depend on the observed nature of the object. For example, it is natural to include the effects of outgassing on an object showing cometary activity. On the other hand, a small apparently asteroidal object is a candidate for the inclusion of solar radiation pressure. In such cases, its detection can yield the bulk density of the object, thereby making it easy to distinguish between the two possible origin cases. 
One such object, known by the internal designation assigned by the initial observer, J002E3, has now been confidently linked to the Apollo 12 fourth stage\footnote{\url{www.jpl.nasa.gov/releases/2002/release\_2002\_178.cfm}}. 
This paper discusses another such object, 2009~BD.

\section{Observational History}

2009~BD was discovered by the Catalina Sky Survey, Mt. Lemmon station, on 2009 January 16 UT \cite{2009MPEC....B...14B}. 
Over the next ten days, the object was observed astrometrically 119 times from 15 different stations. 
A few days later, the apparent magnitude became too faint and the solar elongation too small for continued observation during that apparition. 

The derived visual absolute magnitude was reported as $H=28.8$, which corresponds to a diameter of about $7 \un{m}$ if one assumes a visual geometric albedo of $0.12$, a rough average for asteroids in the inner portion of the solar system. The resulting orbit showed a semimajor axis of $1.01 \un{ua}$, an eccentricity of $0.03$, and an inclination with respect to the mean ecliptic of J2000 of $1.5 \undeg$. Its low $\Delta v$ with respect to the Earth of $3.73 \un{km/s}$ places it at the top of the list of near-Earth asteroids accessible with a spacecraft fly-by\footnote{\url{echo.jpl.nasa.gov/\~{}lance/delta\_v/delta\_v.flyby.html}}.\\

We recovered 2009~BD at Mauna Kea Observatory with two $300 \un{s}$ exposures taken on 2009 October 18 UT using the University of Hawaii $2.24 \un{m}$ telescope, with a confirming observation on 2009 November 1 UT. 
On the latter date, it was necessary to stack eighteen $60 \un{s}$ exposures to reach $R = 22.7$ due to the $98\%$ illuminated Moon being just $12\undeg$ away from the object; background levels reached $44\,000 \un{DN}$ in these unfiltered, two-by-two binned images. Only a single position from the stacked image was reported for this night. 

These original astrometric solutions were performed using the USNO-B1.0 reference catalog \cite{2003AJ....125..984M}, with proper motion for each source taken into account. The inclusion of these new observations by the impact monitoring sites did not eliminate the possibility of a future Earth impact. 
As of this writing, NEODyS shows five possible impacts in the next century\footnote{\url{newton.dm.unipi.it/neodys/index.php?pc=1.1.2\&n=2009BD}}, while Sentry shows eight possible impacts\footnote{\url{neo.jpl.nasa.gov/risk/2009bd.html}}, the earliest being in 2071 in both cases.\\

We observed 2009~BD at a third opposition with two $150 \un{s}$ exposures taken on 2010 July 14 UT, again with the UH $2.24 \un{m}$ telescope on Mauna Kea.
On 2010 September 1 UT we obtained another observation with the $3.6 \un{m}$ Canada-France-Hawaii Telescope on Mauna Kea; the object had magnitude $V\simeq 23.5$, and we obtained a SNR of $6.6$ with a single $300 \un{s}$ exposure.


However, a gravitational orbit solution that incorporated these new observations revealed residuals far in excess of our expected error. Such situations are not uncommon in our experience, and we usually attribute the problem to astrometric reference catalog biases \cite{2005A&A...429..739D}. However, after repeating the solutions using the 2MASS catalog \cite{2006AJ....131.1163S}, which is believed to have substantially smaller biases \cite{2010Icar..210..158C}, the declination bias actually became worse. At this point we began investigating the importance of non-gravitational forces acting on the object.

\section{Methods}

Our initial orbital computation used $119$ astrometric observations collected by various stations in the few days after the discovery, from 2009 January 16 to 26, plus $6$ high-precision positions obtained by us from Mauna Kea in Hawaii, spanning from 2009 October 18 to 2010 September 1.

The first part of the analysis was performed using the radiation pressure option in the software \fo\footnote{\url{www.projectpluto.com/find\_orb.htm}}; it allows the computation of an orbit while adding a seventh free parameter to the usual six orbital elements, namely the Area to Mass Ratio ($AMR$) of the object. The force acting on the object because of the incoming solar radiation is directly proportional to its cross-sectional area exposed to the Sun, and its effect on the orbit (that is, the resulting acceleration) is therefore proportional to the ratio between this cross-section and the mass of the object, that is the $AMR$.
We did not consider the effects of outgassing because there is no obvious cometary appearance in any of our images, the object is too large for the Poynting-Robertson effect to be significant, and electrostatic forces would only be important if there were some mechanism to make the object significantly charged. Because of the small size and likely rapid rotation \cite{2002aste.conf..113P} of 2009~BD, the surface of the asteroid may be sufficiently isothermal to render the Yarkovsky effect undetectable.

The $AMR$ can then be combined with an estimate of the diameter of the object, obtained from its absolute magnitude $H$ and from assumptions on its geometric albedo, to get a range a possible densities. A density value around or above that expected for a porous rock ($\rho \sim 1000 \unrho$) would suggest a natural origin for the object, while a significantly lower value ($\rho \sim 20 \unrho$) would point toward a man-made origin (a hollow metallic shell, such as an upper stage of a rocket).

Both parts of the process ($AMR$ and diameter determination) are based on the assumption that the body is approximately spherical. The implication of a non-spherical shape, together with lightcurve constraints on its importance, are briefly discussed at the end of this work.

\subsection{Outlier rejection techniques}

The computed value for the $AMR$ can be sensitive to each observation included in the solution, and even a few biased astrometric data points may be enough to skew the results, leading to an incorrect evaluation of the importance of radiation pressure on the best-fitting orbit. 

At the same time, the diameter computation relies heavily on the absolute magnitude $H$; however, no calibrated photometry of the object is available, and it is therefore necessary to extract our best estimate of $H$ from the sparse photometry reported by various stations together with their astrometric positions.

Since both of these results are sensitive to the input data, it is important to ensure that the dataset is internally consistent, and all outliers have been rationally rejected, both from the astrometric positions and from the magnitude values.\\

A commonly used method for outlier rejection, known as Chauvenet's criterion \cite{1863QB145.C5.......}, may not applicable in this case, since it is possible that more than one datapoint will need to be rejected, while the criterion is statistically founded only when testing the rejection of a single outlier. Fortunately, another less known rejection method, named Peirce's criterion and first introduced in 1852 by Benjamin Peirce \cite{1852AJ......2..161P}, is statistically correct even in case of multiple rejections (at the cost of a slightly higher mathematical complexity).

For a simplified (but mathematically equivalent) description of the method we refer the reader to \cite{1855AJ......4...81G}, which gives a more astronomy-oriented presentation of the criterion, together with formulas to compute its rejection thresholds in practical cases with a different number of free parameters. Unfortunately the formulas, summarized in \app{Peirce}, are not analytical, and they require an iterative process to obtain a numerical value. However, they can easily be implemented in a computer code, that can give sufficiently accurate values with only a few iterations.\\

The criterion, in its original form, only assumes that the data points are normally distributed around their means, with the same standard deviation\footnote{The last requirement is obviously not satisfied in this case, since every station produces astrometry and photometry with different accuracies. However, we can estimate their typical error bars from other astrometric and photometric data they reported, and then divide the residuals by this expected accuracy; every resulting residual will then be adimensional, and distributed as the standard $N(0,1)$ Gaussian that we need for the rejection criteria. This is equivalent to saying that we will perform our rejection tests in terms of number of sigmas, not in terms of absolute errors.}.
We can therefore directly apply Peirce's criterion to our magnitude rejection problem, since it is safe to assume that the residuals of each magnitude are randomly distributed around the mean, and they reasonably follow a unidimensional Gaussian distribution.\\

Unfortunately, the same is not true for the rejection of astrometric observations; in this case the meaningful metric for accuracy is the angular distance between the observed position and the ephemeris from the fitted orbit, that is the RMS of the right ascension and declination residuals\footnote{We preferred not to use the two coordinates as independent quantities because we wanted to avoid the need of rejecting only one coordinate, and keeping the other. Each astrometric observation is treated as a single entity, and it is either rejected or accepted entirely.}. This quantity, however, is positive definite, and it is obviously not normally distributed around zero; under the assumption that each component of the residual (right ascension and declination) is normally distributed, the RMS is known to follow the Rayleigh distribution (a sort of square root of the chi-square distribution with two degrees of freedom). This distribution can be used to compute a new version of Peirce's criterion that is valid in our case. The mathematical derivation of this result is explained in \app{Peirce}.

\subsection{Photometric debiasing and outlier rejection}

Most of the physical results obtained in this work are sensitive to the assumed value for the absolute magnitude $H$ of 2009~BD. Unfortunately, no accurately calibrated photometry of this object is known to exist, and we are forced to use the poorly calibrated and temporally sparse photometry associated with each astrometric position submitted to the Minor Planet Center (MPC) by various stations.

Such photometric values are known to be affected by various sources of error, both statistical and systematic. It is therefore necessary to perform an accurate analysis of each one of them, to properly estimate its bias with respect to a hypothetical calibrated photometry, its weight in the overall average and the possibility of its complete rejection from the analysis. The complete procedure used to compute our best value of $H$ (in the usual $H$-$G$ formalism from \cite{1989aste.conf..524B}) is quite complex, and it is described in detail in \app{magdeb}.

The application of Peirce's criterion in this case leads to the rejection of photometric data from $3$ nights of the discovery opposition; two of those rejected points, both from station J95, were taken at reasonably high phase angles ($\alpha=52\undeg$ and $\alpha=77\undeg$); this geometry may explain the anomalous offset, either because of a bad assumption of $G=0.15$, or because of the more extreme lightcurve amplitude typical of high phase angle observations \cite{1990A&A...231..548Z}.\\

The remaining photometric values are then used to compute our best value for $H$, and its error bar. We obtain a value of $H=28.43\pm0.12$, where the quoted uncertainty is purely statistical, and does not take into consideration the lightcurve amplitude.

\subsection{Astrometric outlier rejection}

As explained above, the results of this work are also extremely sensitive to the astrometric dataset used as a starting point in the orbital computation, and for this reason we must be sure that all the astrometry is cleaned from biases and systematic effects that can skew the final results. To attain this result, we tried to both maximize the accuracy of the astrometric positions that we produced, and at the same time we used the methods discussed above to reject suspect data points from the astrometry of other stations on which we don't have direct control on the data reduction process.\\

First of all, our astrometric positions from Mauna Kea were all re-reduced against the 2MASS catalog, which is currently considered the least affected by systematic biases in the magnitude range we need. Since 2MASS does not have magnitude information in the visible range, the magnitudes from the USNO-B1.0 reductions were kept and used in the photometric analysis above. Our astrometric and photometric observations are listed in \tab{ours}.\\

We also noticed that a set of eight positions from Tzec Maun Observatory (observatory code H10) showed unusually large residuals; we found that the FITS files of those observations were placed online by the observers, and a total of ten individual exposures were available for download\footnote{From \url{fits.tzecmaun2.org/temp/2009bd\_fits.zip} publicly posted by R. Wodasky on mpml at \url{tech.groups.yahoo.com/group/mpml/message/21487}}; since the object appeared significantly trailed and faint in all those frames, we re-measured them with our astrometry tools, specifically designed to extract high-precision astrometry from trailed detections. A single 2MASS-based position was obtained from a stack of all Tzec Maun exposures, with a total SNR around $25$. We replaced all the other MPC-reported astrometry from code H10 with this single position (also listed in \tab{ours}) and used it in our computations.\\

\begin{table}[htb]
	\centering
	\small
  \tabcolsep 3.5pt
\begin{tabular}{|c|c|c|c|c|}
\hline
      Date [UT] &     $\alpha$ &     $\delta$ &          $R$ &    Code \\
\hline
2009-10-18.429045 & 01 36 20.765 & +04 40 42.69 &       22.0 &        568 \\

2009-10-18.433510 & 01 36 19.469 & +04 40 39.03 &       22.0 &        568 \\

2009-11-01.462459 & 00 50 47.525 & +02 11 14.89 &       22.7 &        568 \\

2010-07-14.587053 & 22 37 44.341 & -22 09 52.01 &       21.4 &        568 \\

2010-07-14.589045 & 22 37 43.649 & -22 09 55.11 &       21.4 &        568 \\

2010-09-01.322573 & 20 07 33.014 & -27 46 14.18 &       23.2 &        568 \\
\hline
2009-01-18.318645 & 07 51 01.037 & +30 03 20.25 &       --   &        H10 \\
\hline
\end{tabular}  
	\caption{Astrometric position and photometry for our 6 observations, plus the stacked re-reduction from the H10 data. The astrometry is referred to 2MASS, while the magnitudes are calibrated against USNO-B1.0.}
	\label{ours}
\end{table}

The second part of our strategy targeted the observations obtained by other stations, all grouped during the first 10 days after the discovery in 2009. Being clustered in a short timespan, they carry less weight compared to ours in the final solution, but it is nevertheless necessary to ensure that at least their average is not biased in any particular direction, which may mimic an offset introduced by radiation pressure effects\footnote{Since a variety of different catalogs were used by the various observatories, and the positions span a range of more than $70 \undeg$ in the sky, we will focus only on rejecting outliers, not on locally debiasing each specific astrometric position.}. 

We applied our ``Rayleigh version'' of Peirce's criterion (discussed in \app{Peirce}) to the full astrometric dataset, including our observations from Mauna Kea.

First of all, for each observation we computed an expected astrometric error bar using the historical performance of the site as published by the Minor Planet Center\footnote{\url{www.minorplanetcenter.org/iau/special/residuals.txt}}. The two-component residuals published by the MPC were combined together in quadrature, to obtain a single error estimate that we assume to have Rayleigh distribution. For our Mauna Kea observations, we used an a-priori error of $0.13\unsec$, an average of the error bars derived during our astrometric solutions\footnote{The error bars derived by the astrometric software range between $0.11\unsec$ and $0.15\unsec$, and are mostly dominated by an assumed $0.10\unsec$ catalog bias. This assumption may be conservative for the 2MASS catalog, but we wanted to make sure that our results were not affected by an excessive weighting of our observations.}. For the remeasured position from station H10 we assumed an error of $0.25\unsec$, from our astrometric software.

This expected accuracy for each station is used twice in the overall orbital computation. First of all, it is converted into a weight in the software \fo, to properly account for each station's performance in the overall orbital fit (in the absence of any better information on the astrometric error). 

We then obtain a first-pass orbit using all available observations, with the seven-parameter radiation-pressure model. This orbit gives us residuals for each astrometric position, and they are then normalized with the error value computed above for each station, as required by Peirce's statistic.

At this step Peirce's criterion is applied (in its Rayleigh form) to reject all possible outliers from the astrometry set. A rejection threshold is computed for the case of $N=118$\footnote{The total number of observations went from $N=125$ to $N=118$ because of the replacement of $8$ positions from H10 with our single stack.} datapoints and $m=7$ fitted parameters, and for various numbers of possible rejections $n$, using the recursive relations in \app{Peirce}. Each value of $n$ is then tested against the data, finding the first value $\bar{n}$ where less than $\bar{n}$ observations have normalized residuals greater than the corresponding computed threshold. In our sample that happens at $\bar{n}=11$, meaning that the $10$ observations with largest normalized residuals need to be rejected from the final orbital solution.\\

This new dataset, now including only $108$ astrometric positions (our $6$ from Mauna Kea, that were all kept by the criterion, plus the $102$ from other stations that survived the rejection process), will be the basis from which our final $AMR$ value will be computed.

\section{Results}

This dataset of $108$ astrometric positions can now be fitted with the software \fo, using the same weighting scheme described above. The best-fitting purely gravitational solution has an RMS of $0.395\unsec$ and a reduced chi-square of $\chi^2_\nu=1.390$, and it shows unreasonably large residuals for some of the Mauna Kea observations (see \tab{Residuals}).

We repeated the orbit solution allowing for solar radiation pressure but otherwise leaving the astrometric data and weights unaltered.
The resulting RMS residual was only $0.291\unsec$, $71\%$ as large as for the gravitation-only orbit, with a significant improvement in the residuals. The reduced chi-square went down by a factor of more than two to $\chi^2_\nu=0.626$, indicating both the significance of the additional parameter and that we were conservative in our estimates of the astrometric uncertainties. The best-fit area to mass ratio associated with this solution is $2.97 \times 10^{-4} \un{m}^2\un{kg}^{-1}$.

The fact that the $\chi^2_\nu$ statistic is not significantly greater than unity indicates that our radiation pressure model is sufficient to satisfy the observations. We therefore do not need to include the Yarkovsky effect, but that does not mean that the effect is not present, just that it remains undetectable in the current dataset.

\begin{table}[htb]
	\centering
	\small
  \tabcolsep 3.5pt
\begin{tabular}{|c|cc|cc|}
\hline
{\bf Date} &   {\bf $\Delta \alpha [\unsec]$} &   {\bf $\Delta \delta [\unsec]$} &   {\bf $\Delta \alpha [\unsec]$} &   {\bf $\Delta \delta [\unsec]$} \\
{\bf [UT]} & \multicolumn{ 2}{|c}{{\bf Grav. only}} & \multicolumn{ 2}{|c|}{{\bf Non-grav.}} \\
\hline
2009-10-18.429045 &     --0.01 &     --0.07 &     --0.03 &     --0.04 \\

2009-10-18.433510 &     --0.02 &     --0.12 &     --0.05 &     --0.09 \\
\hline
2009-11-01.468038 &     --0.19 &      +0.09 &     --0.09 &      +0.17 \\
\hline
2010-07-14.587053 &      +0.26 &     --0.09 &      +0.09 &     --0.01 \\

2010-07-14.589045 &      +0.28 &     --0.09 &      +0.11 &     --0.01 \\
\hline
2010-09-01.322573 &     --0.53 &     --0.13 &     --0.14 &     --0.06 \\
\hline
\end{tabular}  
	\caption{Comparison of the residuals of the purely gravitational solution with those including the radiation pressure effect. Some of the gravitational residuals, such as a $0.53\unsec$ residual for the 2010 September 1 observation, are many sigmas away for our best-estimate for their error bars (typically around $0.13\unsec$ or less for a 2MASS reduction).}
	\label{Residuals}
\end{table}

We then generated $10\,000$ sets of synthetic astrometric observations by adding Gaussian noise to the $108$ available observations. The noise was generated assuming again a standard deviation equal to the RMS of the station, computed from the MPC residuals as explained above. Orbit solutions allowing for solar radiation pressure were performed on each set, yielding an uncertainty in the area to mass ratio of $0.33 \times 10^{-4} \un{m}^2\un{kg}^{-1}$, which represents a $9\sigma$ detection\footnote{The significance quoted here is purely statistical, and it does not take into account possible systematic effects, such as the omission of some other non-gravitational effect.} of that non-gravitational effect.\\

We can now use this estimate of the object's $AMR=(2.97\pm0.33) \times 10^{-4} \un{m}^2\un{kg}^{-1}$, together with our best estimate of its absolute magnitude obtained above ($H=28.43\pm0.12$), to put some constraints on the object's density. 

If we first assume an albedo of $0.12$, the typical average for asteroids in the inner portion of the solar system, we get a density of $\rho \sim 640 \unrho$, which is substantially higher than the densities of the two man-made objects for which similar analyses have been performed and consistent with the density of very porous rock. The Apollo 12 fourth stage (essentially an empty fuel tank) has a density around $20 \unrho$, while the density of the object designated as 6Q0B44E was measured at about $15 \unrho$\footnote{Computed from the $AMR$ estimate available at \url{home.gwi.net/\~{}pluto/mpecs/6q0b44e.htm}, assuming an albedo similar to that used for 2009~BD.}. 

In comparison, the NEAR-Shoemaker spacecraft flyby of (153) Mathilde yielded a bulk density determination of $1300 \unrho$. 
Although such densities are less than what would be expected for solid rock, the fact that Mathilde did not shatter despite the impacts that left craters with diameters a significant fraction of the size of the asteroid strongly suggests an unconsolidated interior structure, with porosity approaching $50\%$ \cite{1999Icar..140....3V}. Furthermore, the object 6R10DB9, now recognized as natural and designated 2006~RH120, has a measured density around $400 \unrho$, comparable with our value for 2009~BD.\\

\subsection{Effect of the unknown albedo}

\begin{figure}[htb]
	\centering
	\resizebox{\hsize}{!}{\includegraphics{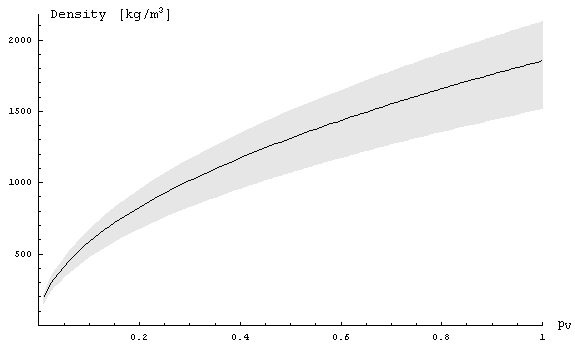}}
	\caption{Estimate of 2009~BD density for albedo values ranging from extremely dark ($p_V=0.01$) to almost complete reflection ($p_V=1.00$).}
	\label{Density}
\end{figure}

Of course, the density of 2009~BD depends strongly on the visual geometric albedo, for which we have no measurement. It may therefore be more meaningful to explore the density values associated with a broad range of albedoes. This analysis is summarized in \fig{Density}, showing that the best-fit density stays well above the typical value for a man-made object down to albedoes smaller than $0.01$.\\

Although the asteroid was near $90\undeg$ solar elongation at about the time that WISE began its all-sky survey in 2010 January, and then again in late June of the same year, the object was not detected. Ephemeris uncertainties were less than $2 \unsec$ at the $1\sigma$ level, so the fact that no WISE source was associated with a known solar system object cannot be attributed to large ephemeris error. 

\section{Conclusions}

The astrometric data reported in this work show the necessity of modeling the orbit of 2009~BD with a perturbation caused by a non-gravitational effect, capable of offsetting the sky-plane position of the object of at least $0.5 \unsec$ during an observed arc of less than 2 years.

We propose that this offset can be explained with the addition of a single parameter, namely an $AMR=(2.97\pm0.33) \times 10^{-4} \un{m}^2\un{kg}^{-1}$, caused by the action of radiation pressure on the object. This result is in agreement with the value $AMR\sim1.9 \times 10^{-4} \un{m}^2\un{kg}^{-1}$ that would be expected for a spherical rocky object of this absolute magnitude ($H=28.43\pm0.12$), assuming water density and an albedo of $0.12$. 

This agreement is consistent with the hypothesis that the effect we are detecting is related to the radiation pressure acceleration. Other non-gravitational effects may contribute on a smaller level. In particular, the multi-parameter Yarkovsky effect is not needed to explain the current dataset. The Poynting-Robertson effect is also negligible for an object of this size, being a factor of $\sim \frac{v}{c}\sim 10^{-4}$ smaller than the radiation pressure effect.

\subsection{Possible origin}

As already mentioned, the density range derived above points toward a natural origin for 2009~BD, excluding the possibility of a man-made object, such as rocket debris, that was left in a heliocentric orbit at the time of the launch. It is therefore necessary to explore other scenarios that may be capable of placing a natural rocky object of this size into an Earth-like orbit.\\

The most direct explanation would be that 2009~BD is a regular NEO coming from the Main Belt through the usual $\nu_6$ resonance mechanism; however, it is unlikely that such a mechanism would produce an orbit resembling that of the Earth, except just by coincidence. Some works, such as \cite{1997Sci...277..197G}, suggest that there may be an enhancement of objects delivered to a low-eccentricity Earth-like orbit with this mechanism, but the overall likelihood of an object being discovered in that state is low (and even lower if we consider that many other objects, such as 1991~VG, 2000~SG344 or the already mentioned 2006~RH120, also share similar Earth-like orbital properties).\\

More likely, this object could be a fragment of the Moon ejected during a collision and cratering process on our satellite. This theory, already proposed to explain the Earth-like orbit of other objects, starting with 1991~VG \cite{1997CeMDA..69..119T}, can directly explain most of the orbital properties of 2009~BD. 

In particular, fragments ejected from the Moon with the lowest possible escape velocity ($v\sim 2.4 \un{km} / \unp{s}$) can enter heliocentric orbits passing through the Earth Lagrangian points; in such cases, the resulting orbit \cite{1995Icar..118..302G} closely resembles that of 2009~BD, with a semimajor axis slightly larger or smaller than the Earth's, eccentricities $< 0.1$ and inclinations $< 1\undeg$. The current inclination of 2009~BD may look larger than this value, but backward integrations show that its eccentricity was well below that threshold until a close encounter with the Earth in 1955. 

The main caveat of this hypothesis is that the size of 2009~BD is close to the upper limit of what can probably be lifted (and achieve escape velocity) from the Moon surface during an impact event. However, studies \cite{1997CeMDA..69..119T} for the case of 1991~VG (an object of similar size, $H \sim 28.4$) confirm that the scenario is plausible and reasonably likely on timescales comparable with the dynamical survival time of an object on such an orbit.

The Moon ejection scenario may be confirmed or disproved with albedo or spectral information, that will hopefully be collected during the next apparition of 2009~BD.\\

Finally, it may be worth mentioning at least two more scenarios, probably much less likely than the one discussed above. 

A direct generalization of the lunar ejection case would be to assume that 2009~BD is a fragment of the Earth surface ejected during a past collision; this possibility is unlikely because the large ejection velocities required to achieve an heliocentric orbit from the Earth's surface would require a massive impact on a timescale less than $10^6 \un{yr}$ ago (the average survival time of objects with this kind of orbits), that would have likely left obvious geological signs on our planet. 

A second hypothesis, that may be worth exploring if the Moon ejection scenario becomes unlikely due to physical observations, is the possibility that the object entered this Earth-like orbit after a very close encounter with our planet, by slowing down significantly by friction during a fly-by phase inside the Earth atmosphere. Similar events have been observed from the ground, and they may lead to Earth-like orbits such as that of 2009~BD.

\subsection{Implications for Impact Monitoring}

2009~BD is currently classified as a possible impactor by both NEODyS and JPL's Sentry, with low-probability impact solutions starting from 2071. The small size of the object, around $8 \un{m}$ assuming an albedo of $0.12$, makes it a negligible threat to the Earth, being at most capable of causing an event similar to the 2008~TC3 impact, with almost no ground damage and only a few meteoritic fragments reaching the ground. The situation can be slightly different if the object is of metallic composition; in that case the bulk of the mass may be able to reach the ground, but it would still be too small to cause any significant and widespread threat or damage. Furthermore, our derived bulk density value clearly points toward a very porous rock, and it is incompatible with a metallic composition, making the scenario of a ground impact unlikely.

However, it is still interesting to explore the implication of the radiation pressure effect on the orbital evolution of the object. The best-fitting gravitational and non-gravitational solutions are now only about $20 \un{km}$ apart, but they will tend to diverge in the future; their distance will grow to about $8\times10^6 \un{km}$, approximately $20$ lunar distances, by the time of the first impact solutions, showing that its effect cannot be neglected in any future impact prediction. 

\subsection{Future Observations}

Our improved orbit for 2009~BD shows that it will pass at $0.9$ lunar distances from the Earth on 2011 June 2, becoming observable as it rapidly increases its solar elongation from $50 \undeg$ on May 28 to $90 \undeg$ on June 1. Peak brightness occurs on 2011 June 3 when the $V$ magnitude reaches $17$.

\begin{figure}[htb]
	\centering
	\resizebox{\hsize}{!}{\includegraphics{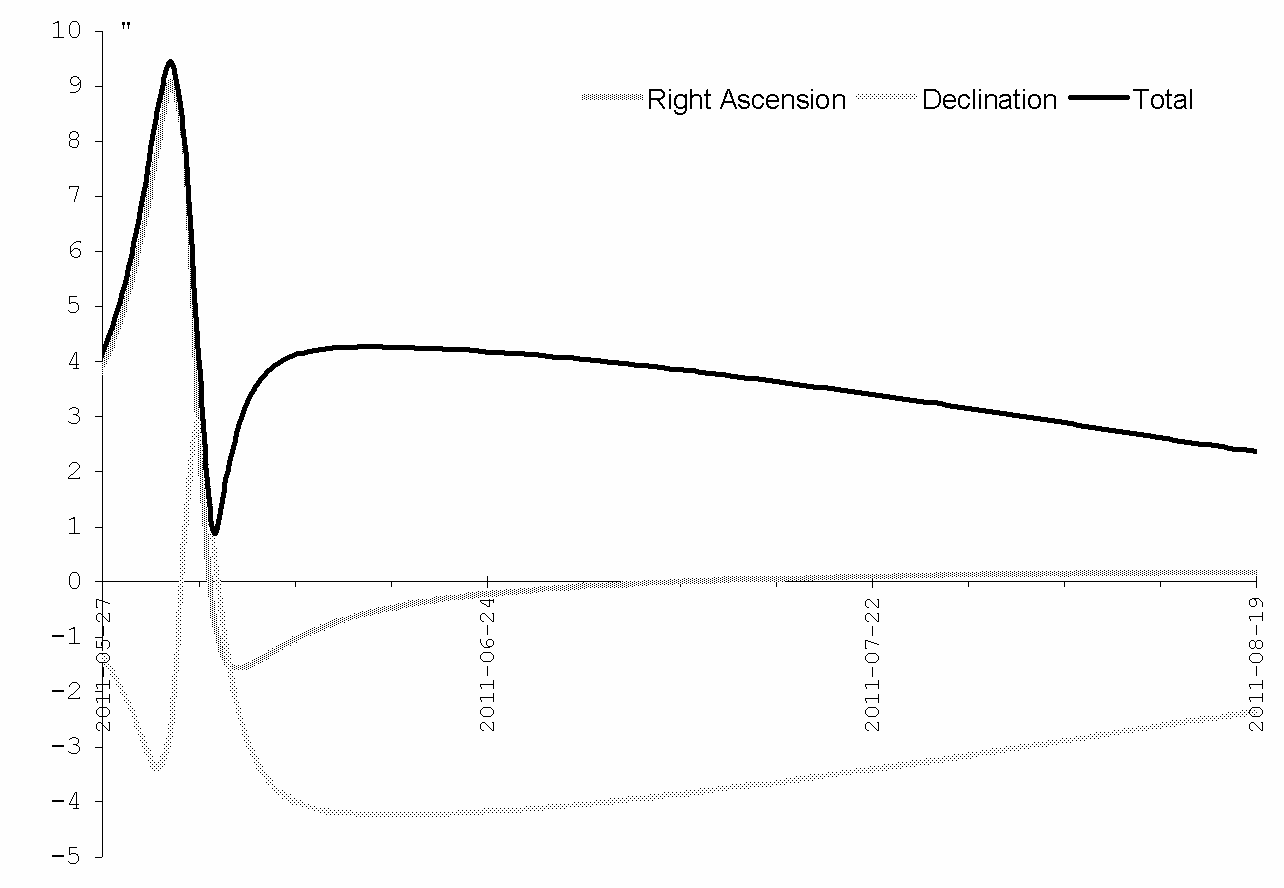}}
	\caption{Offset of the radiation pressure model from a purely gravitational solution, for the 2011 apparition. The plot starts on 2011 May 27, at the time when the object will emerge from solar conjunction.}
	\label{Offset}
\end{figure}

\fig{Offset} shows the ephemeris difference between the gravity-only and gravity-plus-radiation-pressure orbit solutions during this observational opportunity.
The peak difference of $9\unsec$ coincides with the time of close approach, as the slowly increasing linear difference is modulated by the rapidly changing geocentric distance, creating a peak angular difference around perigee.
The right ascension and declination components show different behavior with changing sign that should be easy to detect with careful astrometry.
It should therefore be possible to drastically improve the area to mass ratio determination. 

Other physical observations, including radar, can potentially provide constraints on geometric albedo and size, leading to a good density determination.

\subsection{Acknowledgments}

Our observations of 2009~BD were funded by grant AST 0709500 from the U.S. National Science Foundation.

M.M. wants to thank Valentino Anedda for the help in writing some of the code used in this work.

\appendix 

\section{Peirce's Criterion (and its adaptation to a Rayleigh-distributed quantity)}\label{Peirce}

The following summary of Peirce's criterion is mostly based on the notation introduced in \cite{1855AJ......4...81G}; however, we ignore most of the discussion about the numerical iteration methods, that are not relevant nowadays in light of the current computational capabilities. The computation of the rejection coefficients is necessary in our case, since most precompiled numerical tables (from \cite{1855AJ......4...81G} or other sources) do not extend to the case of $7$ parameter fits, such as the one we need for our orbital computation ($6$ orbital elements plus the $AMR$).\\

The basic idea of Peirce's criterion is similar to Chauvenet's: in a simplified way, we want to identify and reject any datapoint in our sample that lies more than a certain number of ``error bars'' (or sigmas) away from our fitted model, because we consider highly unlikely that such fluctuation may have happened because of random noise. 
The two methods differ only in how this ``certain number'', or rejection threshold, is computed\footnote{And for the already discussed fact that Chauvenet's is statistically valid only for a single rejection, while Peirce's iterative nature makes it usable even for multiple rejections.}.\\

In the basic version from Peirce, where the residuals are assumed normally distributed, the following relations allow us to compute the required rejection threshold, via an iterative process: if we have $N$ observations, of which $n$ show doubtful residuals when fitted with an $m$ parameter model, then the rejection threshold to reject all $n$ observations is at $x$-sigma, where $x$ is the positive solution of:
$$x^2=1+\frac{N-n-m}{n}\left(1-\left(\frac{P_{N,n}}{  R(x) ^n}\right)^{\frac{2}{N-n}}\right)$$
where
$$P_{N,n}=\left(\frac{n}{N-n}\right)^n \left(\frac{N-n}{N}\right)^N$$
$$R(x)=e ^{\frac{x^2-1}{2}} \varphi(x)  $$
and
$$\varphi(x)=\int _x^{\infty }\frac{2}{\sqrt{2\pi }}e^{-\frac{\xi ^2}{2}}d\xi =   \text{Erfc}\left(\frac{x}{\sqrt{2}}\right)$$
\\

As a side note, the rejection threshold for Chauvenet's criterion (let's call it $\chi$, to distinguish it from Peirce's $x$) could  be obtained, in the above formalism, by simply solving the equation
$\frac{1}{2N}=\varphi(\chi)=\text{Erfc}\left(\frac{\chi}{\sqrt{2}}\right)$
\\

With this formalism, a version of the criterion that is valid for our case of positive definite astrometric residuals is easily obtained by replacing the Gaussian function in $\varphi(x)$ with its Rayleigh counterpart: 
$$\varphi'(x)=\int _x^{\infty }2 \xi e^{-\xi ^2}d\xi =   e^{-x ^2}$$
\\
It is interesting to note that, in this case, the function $\varphi'(x)$ is analytically integrable, and it turns out to be a simple exponential function. This fact does not help the rest of the computation of Peirce's threshold, that remains iterative and non-analytical, but it has interesting implications for Chauvenet's criterion: if we solve 
$\frac{1}{2N}=\varphi'(\chi)=\int _\chi^{\infty }2 \xi e^{-\xi ^2}d\xi =   e^{-\chi ^2}$
we obtain an analytical expression for the Chauvenet's threshold of the Rayleigh case, as
$$\chi=\sqrt{\ln{\left(2N\right)}}$$

\section{Debiasing and averaging method to extract the best value for $H$ from sparse MPC photometry}\label{magdeb}

The following text explains in detail the procedure used to extract a value for the absolute magnitude $H$ of 2009~BD from the sparse and uncalibrated photometry made available together with the MPC astrometry. The proposed method deals with debiasing, passband correction, rejection and weighting, and it tries to make the best possible use of all the available data.\\

The procedure starts with an average of all magnitude values of 2009~BD reported by each station in each night. This averaging is done first to reduce the noise in each datapoint, and second to properly take into account cases where a single station reported a large number of photometric points that might otherwise dominate the the final average and torque it toward a possibly biased value. This way, we obtain a total of $28$ photometric datapoints, including $4$ from our own observations from Mauna Kea.

At the same time, the standard deviation of each set of points is also computed, and used as a first guess of the photometric error of the associated datapoint. This ``statistical'' error bar is then added in quadrature to a value of $0.10$ magnitudes; the reason for this correction is twofold: first, it takes into account a ``quantization noise'' introduced by the fact that each photometric point in the Minor Planet Center (MPC) format is given to a $0.1$ magnitude precision, and second it places a lower limit on cases where the ``statistical'' error bar has a formal value of zero (because a single value was used in the average, or all values were equal within the reported precision).\\

Next we need to take into account the different passbands reported by each station, to convert them to a close approximation of the standard $V$ magnitude, that is assumed by definition when computing the $H$ value. At the same time, we need to compensate for known biases in the photometry associated with each station. Both these steps are accomplished during the same process, based on the historical performance of each site, as obtained from the AstDyS website\footnote{\url{hamilton.dm.unipi.it/astdys/index.php?pc=2.0}}.

For each station that reported photometry of 2009~BD, we downloaded from the above mentioned website the complete listing of all their observations of multi-opposition objects. The AstDyS summary file contains an estimate of the bias of each photometric point, computed a-posteriori from the average absolute magnitude of the object. 
These biases take into account a conversion of each passband to $V$, obtained applying a predefined color thought to well represent the average color of an asteroid in that passband. AstDyS uses a correction $V-R=+0.40$ to correct for R-filtered magnitudes, and the same correction is also used for the ``placeholder'' filter code $C$, that was allowed by the MPC at the time of the observations (and stands for ``Clear'', meaning no filter used)\footnote{A correction of $-0.80$ is applied by AstDyS to the cases where a blank is reported in place of a filter code; this is based on the assumption (still in use at the MPC) that a blank filter code stands for a $B$ passband (a standard inherited from the photographic plate time). In most cases, observers reporting ``blank'' magnitudes nowadays mean that no filter was used, and therefore the assumption may be extremely incorrect, introducing an artificial offset in the photometry. However, the method used for our analysis will not be affected by this assumption, since this erroneous artificial offset will show up in the final average bias of the station, and will then be removed by the magnitudes we use in our average for $H$, that were also corrected under the same assumption.}.

We restrict our analysis to those observations obtained in the years 2008 and 2009 (to account for possible changes in the instrumental setting of each observatory), with the same passband and reported by the same program code that measured 2009~BD (when more than one code is present at a station).
We extracted the bias associated to each observations of these multi-opposition objects, averaged together those referred to the same object and the same night, and then averaged all the resulting values to get an estimate of the typical bias of the station.

At the same time, we obtained an estimate of the noise on the bias, measured by the standard deviation of these single-object values. This scatter is mostly due to the rotational lightcurve of each asteroid included in the sample; if we estimate that there is an equal likelihood that each object was observed at any phase of its rotation, we may expect that these effects tend to cancel out in the average, decreasing their influence on the overall bias estimate as $\sqrt{n_s}$, where $n_s$ is the number of observations involved in the average of each station. Therefore, our best guess for the error bar associated to each station-specific bias is the standard deviation of its single biases, divided by $\sqrt{n_s}$; since the $n_s$ of observations involved in the average is quite high for most stations (more than $10^5$ for the surveys, and around $10^2$ for most amateur stations) we obtain error bars on the biases that are usually around $0.05$ magnitudes or less.\\

These values of the station-specific bias are then removed from each photometric datapoint of 2009~BD, and the two error bars (the one from the photometry, and the usually small one from the bias) are combined in quadrature, to get an estimate of the overall quality of the datapoint. These will be the starting values for our computation of $H$.\\

The first pass of the $H$ computation is a weighted fit of these corrected photometric datapoints in the standard $H$-$G$ formalism. Since the quality of the dataset in use is not very high, we are forced to assume a fixed value of $G=0.15$, and restrict our analysis to a single free parameter $H$.
This first fit, computed without rejections, gave a best-fitting value of $H=28.40$. Residuals of each photometric point with respect to this value were then computed, and they formed the basis for the rejection process, using Peirce's criterion.\\

The application of the criterion to the magnitude dataset is straightforward from a mathematical point of view, since these residuals can be assumed to be normally distributed.
However, a further physical complication is introduced by the fact that each residual is composed by two different parts, one given by the statistical noise, and one by the intrinsic lightcurve of the object, of unknown amplitude, period and phase. If we decide to reject a point based on the fact that its residual is a given number of sigmas away from zero, this may be due to the point being located at a maximum (or minimum) of the lightcurve, and not to its low quality. Because of the lightcurve, even a very high-quality photometric point may be rejected by a straightforward application of the rejection criterion, if it lays at a lightcurve extreme.

A first way to partially address this problem would be to artificially inflate the error bars of each data point, adding the lightcurve amplitude to each photometric error. However, this would require an independent knowledge of the lightcurve amplitude, which is not available in this case\footnote{Some uncalibrated photometry of 2009~BD, obtained from station 095, was made available online at \url{www.astroalert.su/2009/01/18/neo2009bd}; it shows that the object may have a lightcurve variation of approximately $1$ magnitude, and it seems to be variable on a timescale of 10 minutes or less. A clear evidence for lightcurve variation on the same timescale, without a period and an amplitude, can also be inferred from the data from station H10 mentioned in the text.}.

For this reason, we have to use an alternative, indirect method to extract the same information from the available data, in a statistical sense.
If we assume that the residuals, after the first-pass fit of $H$, are due to the sum of the two effects (statistical error and lightcurve), we may expect that they are, on average, the sum (in quadrature) of these two components. We may then try to recover the unknown one of these components (the lightcurve amplitude, $\Delta_l$) from the other component (the statistical error, $\Delta_s$) and the total error (measured by the residuals). If we do so on the average of all points, we see that the average statistical error is around $\bar{\Delta_s}\sim0.16$ magnitudes, while the residuals are scattered with $RMS\sim0.41$ magnitudes. If we assume, in symbols, that ${\bar{\Delta_s}}^2+{\Delta_l}^2=RMS^2$, then we get $\Delta_l\sim0.38$, our estimate of the lightcurve amplitude\footnote{This lightcurve amplitude is estimated in an ``RMS'' sense, and does not correspond to the peak-to-peak amplitude usually quoted for other objects. A conversion between the two may be obtained only under the assumption of a functional form for the lightcurve itself: if we consider a purely sinusoidal case, then the peak-to-peak amplitude is equal to $2\sqrt{2}\simeq 2.83$ times the RMS value, or about $1.07$ magnitudes in our case, in good agreement with the estimate from station 095 mentioned above.}.\\

We can now add this estimate of the lightcurve amplitude to the error of each data point, and use this new error bar to estimate how many sigma each residual is away from the average. This is what we need to perform a rejection with Pierce's criterion, as discussed above. The test rejects $n=3$ photometric datapoints, and the remaining $25$ values are then used to recompute a final value for $H$, and its error bar.\\

\bibliographystyle{my}
\bibliography{Articles}

\end{document}